\documentclass[aps,prb,twocolumn,showpacs,amsmath,amssymb,superscriptaddress]{revtex4-2}
\usepackage{graphicx}
\usepackage{color}
\usepackage[colorlinks,bookmarks=false,citecolor=blue,linkcolor=red,urlcolor=blue]{hyperref}
\usepackage{multirow}
\usepackage{ulem}
\usepackage{tabularx}
\usepackage[nounderscore]{syntax}
\graphicspath{{fig/}}

\newcommand{\be}{\begin{equation}}
\newcommand{\ee}{\end{equation}}

\begin{document}
\title{Exotic surface behaviors induced by geometrical settings of the two-dimensional dimerized quantum XXZ model}
\author{WenJing Zhu}
\affiliation{Department of Physics, Beijing Normal University, Beijing 100875, China} 
\author{Chengxiang Ding}
\email{dingcx@ahut.edu.cn}
\affiliation{School of Science and Engineering of Mathematics and Physics, Anhui University of Technology, Maanshan, Anhui 243002, China}
\author{Long Zhang}
\email{longzhang@ucas.ac.cn}
\affiliation{Kavli Institute for Theoretical Sciences and CASD Center for Excellence in Topological Quantum Computation, University of Chinese Academy of Sciences, Beijing 100190,
China}
\author{Wenan Guo}
\email{waguo@bnu.edu.cn}
\affiliation{Department of Physics, Beijing Normal University, Beijing 100875, China}
\affiliation{Beijing Computational Science Research Center, Beijing 100193, China}
\date{\today}

\begin{abstract}
    We study the surface behavior of the two-dimensional columnar dimerized quantum antiferromagnetic XXZ model with easy-plane anisotropy, with particular emphasis 
    on  
    the surface critical behaviors of the (2+1)-dimensional quantum critical points of the model that belong to the classical 
    three-dimensional O(2) universality class, 
    for both $S=1/2$ and $S=1$ spins using quantum Monte Carlo simulations.
    We find completely different surface behaviors on two different surfaces of geometrical 
    settings: the dangling-ladder 
    surface, which is exposed by cutting a row of weak bonds, and the dangling-chain surface, which is formed by cutting a row of strong bonds along 
    the direction perpendicular to the strong bonds of a periodic system. 
    Similar to the Heisenberg limit, we find an ordinary transition on the dangling-ladder surface for both $S=1$ and $S=1/2$ spin systems. 
However, the dangling-chain surface shows much richer
    surface behaviors than in the Heisenberg limit. For the $S=1/2$ easy-plane model, 
at the bulk critical point, we provide evidence supporting an extraordinary surface transition with a long-range order established 
   by effective long-range interactions due to bulk critical fluctuations. The 
   possibility that the state is an extraordinary-log state seems unlikely. For the 
   $S=1$ system, we find surface behaviors similar to that of the three-dimensional 
   classical XY model with sufficiently enhanced surface coupling, suggesting an extraordinary-log state at the bulk critical point.
    
\end{abstract}

\maketitle

\section{Introduction}

At bulk critical points, surface states may emerge with 
novel critical phenomena beyond the bulk critical properties \cite{binder, BH, Diehl, Die97, Deng}. 
Typically, there are three classes of such surface states, namely, ordinary, special, and extraordinary, e.g., the three surface states of the classical O($n$) model in
dimension $d>3$ at its bulk critical point.  The fixed point of the extraordinary state corresponds to the bulk transition occurring in the presence of an ordered 
surface. This can happen if surface couplings are strong enough, such 
that the surface orders before the bulk when the temperature is lowered.
On the other hand, if the surface couplings are the same in the bulk, one expects an ordinary transition, where singularities of the state come entirely 
from the bulk criticality.
The special point is a multicritical point separating the ordinary and extraordinary states. 
This picture is supported by renormalization group theory \cite{Diehl, Die97}. 

However, due to the Mermin-Wagner theorem \cite{Mermin}, it is widely accepted that 
there is no extraordinary state, thus, there is no special state, for the $n\ge 2$ O($n$) model
in dimension $d=3$, since the two-dimensional (2D) surface of the model cannot hold
long-range order. Recently, Metlitski proposed a field-theoretical study to 
investigate the classical O($n$) model with $n$ being continuously varied and suggested a 
new extraordinary-log class of 
the surface states for $n=2$ and $n=3$, where the surface correlation function decays
as a power of $\log(r)$ \cite{ContinuousN}. Some predictions in this work have 
been verified in 
classical three-dimensional (3D) O(3) and classical 3D O(2)
models. \cite{Toldin,lvjp}

For unfrustrated quantum systems, 
the $d-$dimensional SU(2) quantum critical point can be described by the $(d+1)-$
dimensional classical O(3) universality class, according to the quantum to classical 
mapping. Therefore, similar surface critical behaviors (SCBs) are expected for $d = 2$ quantum critical points in the 3D O(3) universality class. However, some 
unexpected behaviors were found for the surface transitions of the quantum critical points.

For dimerized antiferromagnetic Heisenberg systems, where the 
couplings are not distributed homogeneously on the lattice
and are directionally anisotropic, thus making it possible to expose different surfaces 
by geometrical settings, 
one finds different SCBs on different surfaces  without adapting surface couplings. 
\cite{zhanglong, Ding2018, Weber1} 
In particular, on the surface formed by dangling spins, nonordinary SCBs are 
found. The mechanism 
of such behavior is not yet clear. It was attributed to the surface capturing the 
gapless state in the bulk gapped phase due to
topological $\theta$-term of the $S=1/2$ antiferromagnetic dangling spin chain. 
However, later research on the $S=1$ dimerized Heisenberg antiferromagnet showed
similar nonordinary SCBs on the dangling-spin surface, in which 
the topological-$\theta$ term is absent and the surface is thus gapped in the bulk disordered phase\cite{Weber2}. This has led to the above scenario being doubted.

The symmetry-protected topological (SPT) state\cite{ZCGu, Pollmann} offers another possible mechanism for 
such nonordinary SCBs of quantum criticality in the 3D O(3) 
universality class \cite{zhanglong,CHC}.  In a recent paper\cite{CHC}, we showed
that the surface 
perpendicular to the coupled topological $S=1$ Haldane chains displays 
nonordinary SCBs, which is understood as the ends of the chains forming an $S=1/2$
chain, which is gapless through the bulk gapped phase and thus forces the system to
enter a nonordinary surface critical state.

To date, for the SCBs of the 3D O(2) universality class, only 
classical models have been studied \cite{Deng,lvjp}. The SCBs of the quantum criticality belonging to 
the 3D O(2) universality class, especially with the quantum nature of
spins, i.e., $S=1/2$ and $S=1$ involved, have not been investigated.
In consideration of the new and interesting phenomena that have been observed and the mechanisms that are 
under debate in quantum models of the 3D O(3) universality class, it would be beneficial 
to study quantum models hosting the quantum criticality in the 3D O(2) universality class.
In this work, we study the surface critical behaviors of 2D columnar
dimerized (CD) spin-$S$ quantum antiferromagnetic (QAF) XXZ models with 
easy-plane anisotropy that host quantum phase transitions from the N\'eel state to the 
quantum dimer state, which belongs to the classical 3D O(2) universality class, 
using unbiased quantum Monte Carlo simulations. 
Similar to the dimerized antiferromagnetic Heisenberg model studied previously, 
which is the isotropic limit of the current model, 
we form and study dangling-ladder and dangling-chain surfaces.

The dangling-ladder surface (to be specified in Sec. \ref{model} ) is gapped 
in the bulk gapped phase for both
$S=1/2$ and $S=1$. Without tuning the surface couplings, it is natural to 
expect ordinary SCBs on this surface. Indeed, we observe ordinary 
SCBs in the dangling-ladder surface for both $S=1/2$ and $S=1$ spins. 

However, we observe different surface behaviors in the dangling-chain surface (to be
specified in Sec. \ref{model}) for the $S=1/2$ and $S=1$ models.

On this surface, the $S=1$ model behaves similarly to 
the classical 3D O(2) model with large surface couplings, where we find
a transition in the bulk gapped phase, which can be understood as a
Berenzinsky-Kosterlitz-Thouless (BKT)
transition \cite{Berezinsky1970,Berezinsky1972,Kosterlitz}. Further, the surface SCBs 
seem to belong to the extraordinary-log fixed point \cite{ContinuousN}
at the bulk critical point.
 
For the $S=1/2$ model, we find unexpected behaviors, e.g., divergence of the scale surface correlation length $\xi_1/L$ 
preceding the bulk critical point, and crossings of $\xi_1/L$ for different system sizes
in the bulk gapped phase, considering that the surface is already critical
as a Tomonaga-Luttinger liquid (TLL) \cite{Tomonaga,Luttinger} in the dimerized limit. 
One scenario to understand this is that there is a new surface state,
between the transition point and the bulk critical point,
with spin-spin correlation decaying as a power of $\log(r)$, different from the TLL, and with a finite correlation length
to the bulk. 
We do not have a clear understanding of this phase and 
the corresponding phase transition. However, a transition from two 
different two phases without long-range order is not excluded by the
Mermin-Wagner theorem \cite{Mermin}. Such transitions have been found in 
2D O(2) \cite{Domany} and 2D O(3) \cite{WAGuo} systems. Furthermore, as found in classical models, 
a changing in the interaction range, 
not necessarily infinite, can drive the phase to another fixed point\cite{xfqian}.
However, although the data can be fit according to the proposed scaling formula, the fits are not stable upon gradually 
excluding small systems. It is more likely
that the correlation behaves finally cross over to normal TLL behavior. 
In the end, it is more plausible that the divergence of $\xi_1/L$ are results of bulk critical behavior.

At the bulk critical point, we further study 
the SCBs of the dangling-chain edge. 
Our numerical results support an extraordinary SCB with long-range
order established by effective long-range interactions due to bulk
critical correlations. The possibility that the surface shows 
extraordinary-log behaviors seems unlikely.

This paper is organized as follows. 
Sec. \ref{model} describes the models and methods. The observables used to
investigate the surface states are also defined.
In Sec. \ref{Dladder}, we present the results of the SCBs on 
the dangling-ladder surface.
Sec. \ref{DChain} describes the studies of the SCBs on the dangling 
chain surfaces.
In Sec. \ref{conclusion}, we conclude the paper. 

\section{Models and Method}
\label{model}
\begin{figure}[!ht]
      \includegraphics[width=1\columnwidth]{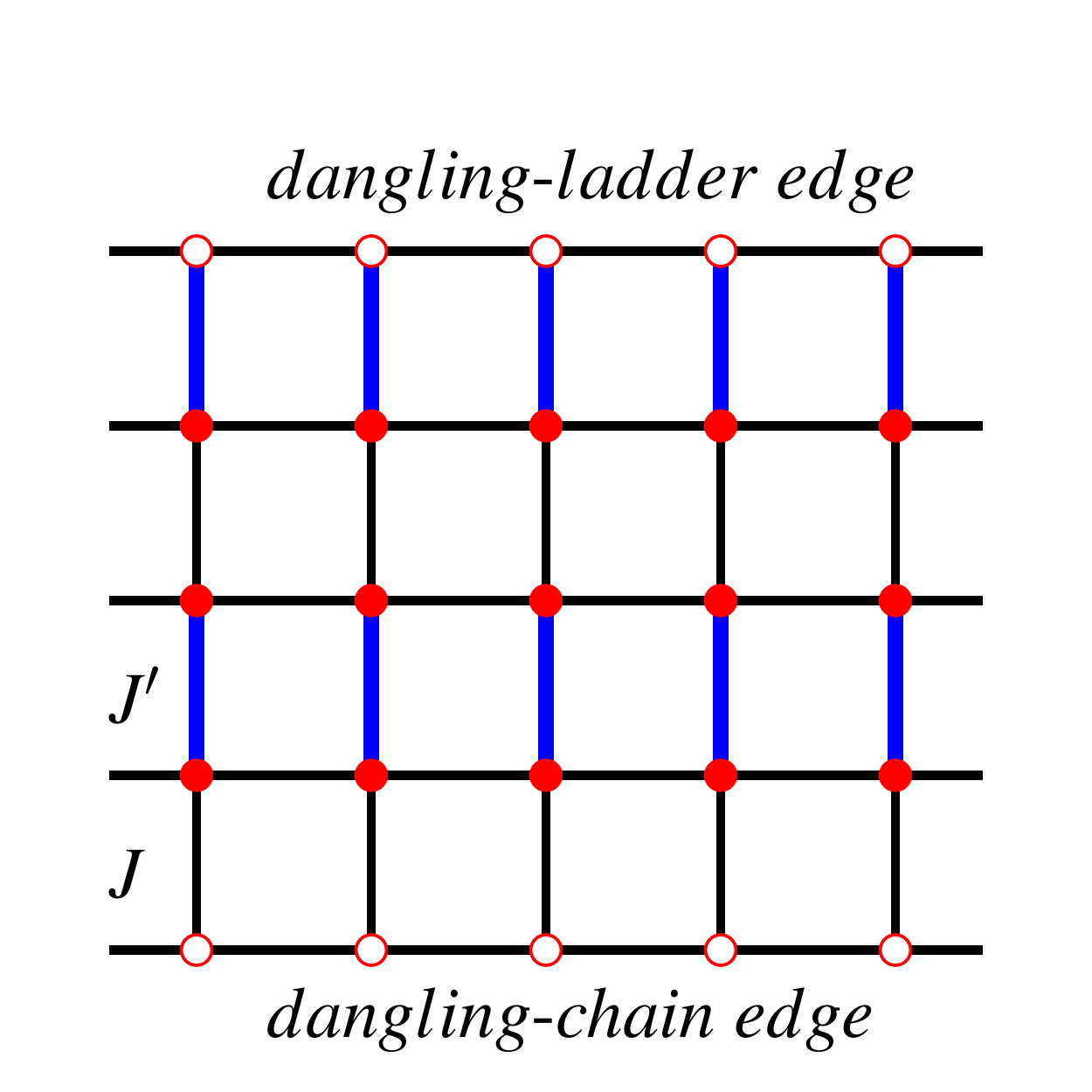}
    \caption{
        Columnar dimerized easy-plane QAF XXZ model with open boundaries indicated by open red circles. \(J(J^\prime)\) are 
        antiferromagnetic XXZ couplings denoted by black (blue) bonds. 
        By maintaining periodic boundary conditions in the $x$ direction, the model exposes a dangling-chain edge (below) and a dangling-ladder edge (upper) by cutting \(J^\prime\) and \(J\) bonds, respectively. 
        \label{fig:CDlat}
    }
\end{figure}

We study the columnar dimerized spin-$S$ quantum antiferromagnetic XXZ model on the 
2D square lattice, as shown in Fig. \ref{fig:CDlat}. 
The Hamiltonian of the model is given by
\begin{eqnarray}
H = J\sum_{<i,j>} D_{ij} + J^\prime \sum_{<i,j>^\prime} D_{ij},
\end{eqnarray}
where $J$ and $J^\prime$ represent weak and strong antiferromagnetic couplings, denoted by black and blue bonds, respectively;
$D_{ij} = S_i^x S_j^x + S_i^y S_j^y + \Delta S_i^z S_j^z$, with $\Delta$ the anisotropy parameter. 
We set \(J=1\) and let \(g=J^\prime/J>1 \) be the tuning parameter driving 
phase transition. 

By tuning the anisotropy parameter \(\Delta\) away from $1$, the system 
breaks the SU(2) symmetry down to the U(1) symmetry in easy-plane 
with \(\Delta \in [0,1)\) or \(Z_2\) symmetry in easy-axis $z$ with 
\(\Delta > 1\). 
The system undergoes a conventional Wilson-Fisher transition from the
N\'eel state breaking the U(1) symmetry to the dimerized state in the 
3D O(2) universality class by tuning \(g\) in the easy-plane case or 
a transition from the N\'eel state breaking the $Z_2$ symmetry to the dimerized
state in the 3D Ising universality class in the easy-axis case. 
Here, we focus on the $S=1/2$ and $S=1$ easy-plane models. 

In this work, we adopt the stochastic series expansion (SSE) \cite{SSE} 
quantum Monte Carlo (QMC) method with loop algorithms \cite{directedloop,loop2} to 
simulate the above models. 
We have specified \(L_x = L_y = L\), 
where \(L_x\) and \(L_y\) are linear lattice sizes along $x$ and $y$ 
direction, respectively. The inverse temperature \(\beta\) scales as $2L$ 
to calculate the bulk critical points and as $L$ to study the SCBs of the
models, respectively, considering the dynamical critical exponent $z=1$ for 
the bulk criticality. Typically, $10^7$ Monte Carlo samples are taken to 
calculate the average and estimate each data point's errorbar.

The bulk critical properties for a given \(\Delta\) and $S$ are obtained 
by finite-size scaling analyses on 
the spin stiffness obtained by quantum Monte Carlo simulations. 
The results are listed in Table \ref{tab:cri}. More details are presented in Appendix \ref{cri_bulk}.

\begin{table}[!ht] 
     \caption{Bulk quantum critical points (QCPs) and exponents $\nu$ for the columnar dimerized easy-plane QAF XXZ models of given \(\Delta\) and $S$. We also list properties published in the literature for the reader's convenience. 
     }
    \begin{tabular*}{\columnwidth}{c @{\extracolsep{\fill}} cclll}
        \toprule
           Type   & S &$\Delta$&   $g_c$                     &    $\nu$      \\
           \hline
        \multirow{5}{*}{CD}
            & 1/2 & 0.5 & 2.73227(5)                     & 0.67(1)     \\
            & 1/2 & 0.9 & 2.1035(1)                   & 0.672(4)     \\
            & 1   & 0.9 & 6.01(1)                       & 0.66(1)      \\
            & 1/2 & 1   & 1.90951(1) \cite{yao2018}    &       \\
            & 1   & 1   & 5.2854(6) \cite{Weber2}      &              \\
    \hline
      \toprule
    \end{tabular*}
    \label{tab:cri}
\end{table}

To investigate the SCBs, 
we apply periodic boundary conditions in the $x$ direction 
and expose open boundaries by cutting \(J \) or \(J^\prime \) bonds 
along the $x$ direction.
The open boundaries are indicated by open red circles in Fig. \ref{fig:CDlat}. 
Note that the boundary exposed by cutting \(J\) bonds exhibits a gapped 
ladder in the limit $g \to \infty$. Hence, this boundary is referred 
to as the dangling-ladder surface. 
The edge exposed by cutting a row of \(J^\prime \) bonds is referred to as the 
dangling-chain edge, as it behaves as a quantum chain in the large $g$ limit. 

We adopt the squared surface staggered magnetization \(m_{s1}^2\) 
and the surface spin-spin correlations to characterize the surface critical behaviors. 
These quantities depend on two equal-time spin-spin correlations defined as 
\be
C^{\rm XY}(i, j) = \langle (S_i^x S_j^x +S_i^y S_j^y)\rangle
\ee
and
\be
C^{\rm Z}(i,j) = \langle S_i^z S_j^z \rangle.
\ee
While $C^{\rm Z}(i,j)$ can be directly estimated in the $S^z$ representation, the in-plane correlation $C^{\rm XY}(i,j)$ is calculated through the following Green function
\be
G(i,j)= \frac{1}{2} (\langle S^+_i S^-_j\rangle +\langle S^-_i S^+_j\rangle),
\label{GF}
\ee
which is determined efficiently with an improved estimator by using the
loop updating algorithm in the SSE QMC simulations.\cite{Evertz, Troyer}

The surface parallel correlation
$C_\parallel(L/2)$ averages $C^{\rm XY} (i,j)$ between 
two surface spins $i$ and $j$ at the longest distance $L/2$ over the 
surface under consideration, and the perpendicular correlation function
$C_\perp(L/2)$ averages  $C^{\rm XY} (i,j)$ between 
two spins $i$ and $j$ at the longest distance $L/2$, where $i$ is fixed on 
the surface and $j$ is located at the center of the bulk, with the direction 
from $i$ to $j$ perpendicular to the surface. 

The squared surface staggered magnetization is 
defined as
\begin{equation}
m_{s1}^2  = 
\frac{1}{L^2} \langle ( \sum_{i \in {\rm surface}} (-1)^{i} S^x_i)^2 +  ( \sum_{i \in {\rm surface}} (-1)^{i} S^y_i)^2 \rangle,
\end{equation}
where $i$ labels the spins on the surface, which can be expressed with 
the Green function defined in Eq. (\ref{GF}),
\be
m_{s1}^2= \frac {2} {L^2} \sum_{i,j \in {\rm surface}} (-1)^{i+j} 
G(i,j)
\label{eqn:mxy}
\ee
where $i, j$ are the spins on the surface. 

In addition, we calculate the surface second-moment correlation length
\begin{equation}
\xi_1 = \frac {L} {2\pi} \sqrt{\frac {S_1(q)} {S_1(q+\Delta q)} -1 },
\end{equation}
with  $q=\pi$ and $\Delta q=2\pi/L$, 
where the surface spin structure factor $S_1(q)$ is the Fourier transform of the surface spin-spin correlation 
$C_\parallel^{\rm XY}(r_{ij})$ which averages $C^{XY}(i,j)$ between 
two surface spins $i$ and $j$ at the distance $r_{ij}=|j-i|$
\begin{eqnarray}
S_1(q) =\frac{1}{\sqrt{L}}\sum_{i,j \in {\rm surface}} e^{-i q (j-i)} C^{\rm XY}_\parallel (r_{ij}).
\end{eqnarray}

\section{surface critical behaviors of the dangling-ladder edge}
\label{Dladder}

In this section, we study the dangling-ladder edge of the $S=1/2$ and $S=1$ 
models. This edge is formed by  ``nondangling" spins, see 
Fig. \ref{fig:CDlat} (the top edge).  It is also called nondangling surface.

For the $S=1/2$ quantum spin models with SU(2) symmetry, the bulk 
transition is 
in the classical 3D O(3) universality class. \cite{Matsumoto}
For the dangling-ladder edge, the surface critical exponents are 
found in agreement with the ordinary universality class \cite{Ding2018,Weber1}. 
This surface state is understandable since the edge can be viewed as 
a dangling ladder in the gapped dimer phase, and the ladder is gapped since it has 
two legs, thus effectively forming an integer-spin chain. 
It is different from the edge formed by 
dangling spins, as shown in Fig. \ref{fig:CDlat} (the bottom edge), which 
is gapless in the gapped dimer phase, as a result, the surface critical behavior 
is nonordinary \cite{Ding2018, Weber1}. 
When the symmetry of the model is down to $U(1)$, the gapped nature of the 
dangling ladder does not change. Therefore, we expect ordinary SCBs of 3D O(2) class 
on the edge. Our numerical results, as shown below, confirm this.

For the $S=1$ critical system in the 3D O(3) universality class, the
dangling-ladder edge is gapped as well. 
It was found that the SCBs are the ordinary transition of the 3D O(3) universality 
class\cite{Weber2}. 
When the symmetry is down to U(1), ordinary SCBs of the 3D O(2) universality class are 
expected and verified by our numerical results discussed in this section.

We simulated $S=1/2$ systems with linear sizes up to $L=256$ and 
$S=1$ systems with linear size up to $L=96$. 
The quantities $\xi_1/L, C_\parallel(L/2), C_\perp(L/2)$ and $m_{s1}^2L$ were 
calculated. 

\begin{figure}
    \includegraphics[width=1\columnwidth]{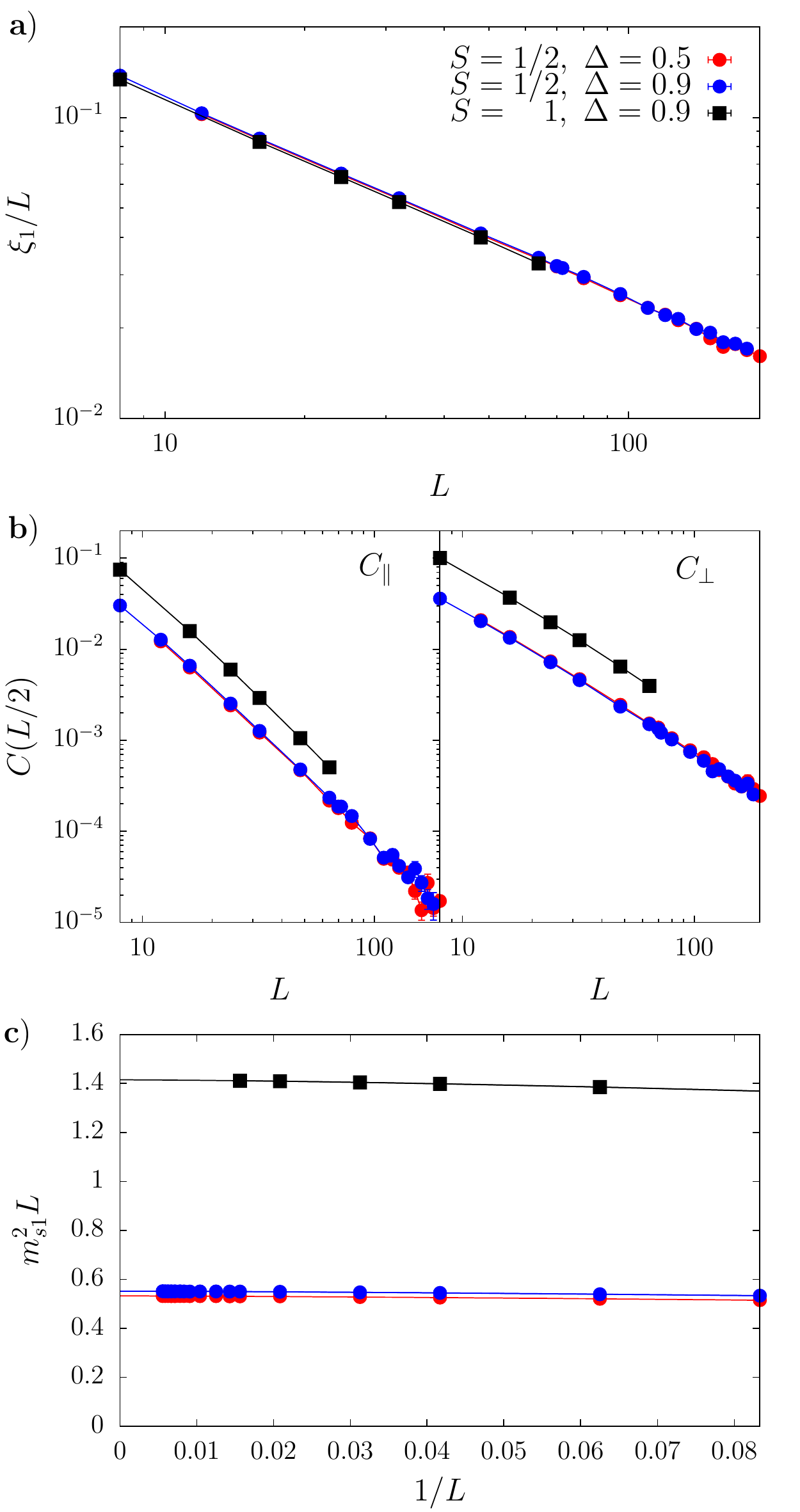}
    \caption{
        Surface quantities on the dangling-ladder edge for the CD easy-plane XXZ model at the bulk QCPs with $\Delta = 0.5$ and $\ 0.9$ for the $S=1/2$ case and $\Delta = 0.9$ for the $S=1$ case. 
        a) The scaled surface correlation length $\xi_1/L$ as a function of $L$ on a log-log scale. 
        b) The surface parallel correlation function $C_\parallel(L/2)$ and the perpendicular correlation function $C_\perp(L/2)$ as a function of $L$ on a log-log 
	scale. 
        c) The scaled surface magnetization $m^2_{s1}L$ vs $1/L$.
    }
	\label{cut-J}
\end{figure}

In Fig. \ref{cut-J} (a), we show the scaled surface correlation length 
$\xi_1/L$ as a function of the system size $L$ on a log-log scale. It is clear 
that the quantity converges to zero algebraically.  
By fitting data to the following formula\cite{zhanglong},
\be
\xi_1/L \propto L^{-p}, 
\ee
we find $p=0.681(5)$ for the $\Delta=0.5$ $S=1/2$ model, $p=0.69(1)$ for the $\Delta=0.9$ $S=1/2$ model, and $p=0.67(2)$ for the $\Delta=0.9$ $S=1$ model. 
The decay of $\xi_1/L $ to zero shows that the surface correlation
length diverges more slowly than the bulk correlation length $\xi$. Further, the 
surface critical behavior is controlled by $\xi$.

Figure \ref{cut-J} (b) shows the parallel correlation $C_\parallel(L/2)$ and perpendicular correlation $C_\perp(L/2)$ as functions of the system size $L$ on a log-log 
scale. Both decay as power laws. We expect the following finite-size scaling behavior 
for them \cite{zhanglong,Ding2018,Weber1}:
\be
C_\parallel(L/2)=L^{-(d+z-2+\eta_\parallel)}(a +b L^{-\omega}),
\label{cpara}
\ee
and
\be
C_\perp(L/2)=L^{-(d+z-2+\eta_\perp)}(a +b L^{-\omega}),
\label{cperp}
\ee
where $\eta_\parallel$ and $\eta_\perp$ are two surface anomalous 
dimensions, $\omega >0$ represents
the effective exponent controlling corrections to scaling, which  
differs in the above two equations; $d+z$ is the space-time
dimension with $d=2$ and $z=1$ in the current models.

The squared surface staggered magnetization $m_{s1}^2(L)$ obeys the
following finite-size scaling form\cite{zhanglong,Ding2018} for $d=2$ and  $z=1$:
\be
m_{s1}^2 L = c + L^{2y_{h1} - 3} (a +b L^{-\omega}),
\label{ms1fs}
\ee
in which $y_{h1}$ is the scaling dimension of the surface staggered magnetic field $h_1$; $c$ is a nonuniversal constant representing analytic correction 
to scaling due to surface. As shown in Fig. \ref{cut-J}(c), $m^2_{s1}L$ converges
to a constant $c$ as $L \to \infty$ for all three cases studied.

By gradually excluding 
small system sizes, we obtain statistically sound fits of Eq. (\ref{cpara}), Eq. (\ref{cperp}) and Eq. (\ref{ms1fs}) without corrections to scaling 
for the $S=1/2$,
$\Delta=0.5$ and $0.9$ systems, and for the $S=1$ and $\Delta=0.9$ systems, respectively. 

For the $S=1/2$ spins, in the case $\Delta=0.5$,
we find $\eta_\parallel = 1.34(2)$, $\eta_\perp = 0.644(6)$ ,  and $y_{h1} = 0.82(1)$;
for $\Delta=0.9$, we obtain $\eta_\parallel = 1.41(2) $ and $\eta_\perp = 0.69 (2)$,  and $y_{h1} = 0.80(1) $. 
The exponents for the two different $\Delta$ are consistent within 
2 error bars. 
For the $S=1$ case, we find exponents $\eta_\parallel=1.51(1)$, 
$\eta_\perp=0.65(1)$, and $y_{h1}=0.74(5)$ for $\Delta=0.9$. 
These results are listed also in Table \ref{tab:ord}.

The exponents $y_{h1}$ estimated for these models are close to those found for 
the ordinary 
SCB in the classical 3D O(2) models \cite{NB,LPB,Deng}. 

The surface critical exponents are expected to satisfy the following 
scaling relations \cite{BH,Barber,Lubensky}
\be
2 \eta_\perp = \eta_\parallel + \eta
\label{scaling1}
\ee
and 
\be
\eta_\parallel = d+z - 2 y_{h1}
\label{scaling2}
\ee
with $\eta$ being the anomalous magnetic scaling dimension of the bulk 
transition in the $d+z$ spacetime. 
For the easy-plane XXZ model with 
U(1) symmetry in two dimensions, we have $d+z =3$ and $\eta=0.038$\cite{xy}. 
The exponents found for SCBs of the easy plane XXZ models with variant
$\Delta$ and $S$ satisfy or nearly satisfy the above scaling relations within error bars. The relatively large deviation for the $S=1$ $\Delta=0.9$ model might 
be
 because the simulated system sizes are not large enough.

\begin{table}[!ht]
     \caption{SCB exponents for the dangling-ladder
     edge of the CD easy-plane XXZ models. The exponents $y_{h1}$ of the three
     models are close to $y_{h1}=0.781(2)$ found for the ordinary surface transition of the classical 3D O(2) model\cite{Deng}.
     }
    \begin{tabular*}{\columnwidth}{c @{\extracolsep{\fill}} cclll}
        \toprule
           Type   & S &$\Delta$&   $\eta_\parallel$                     &    $\eta_\perp$     &$y_{h1}$ \\
           \hline
        \multirow{3}{*}{CD}
            &1/2  &0.5 &       1.34(2)              & 0.644(6)    & 0.82(1) \\
            &1/2  &0.9 &       1.41(2)             & 0.69(2)     & 0.80(3)  \\
            & 1   & 0.9 &      1.51(1)             & 0.65(1)    &  0.74(5) \\
    \hline
      \toprule
    \end{tabular*}
    \label{tab:ord}
\end{table}

\section{Surface critical behaviors of dangling-chain edge}
\label{DChain}

We now consider the dangling-chain edge for the CD QAF easy-plane XXZ 
model with \(0 \le \Delta \le 1\). 

For the SU(2) CD models, the dangling-chain edge undergoes
nonordinary SCBs with exponents close to the special transition of the 3D O(3) 
universality class\cite{Ding2018,Weber1}. This was attributed to the fact that  
the edge is gapless due to the topological $\theta$ term of the $S=1/2$ dangling chain 
in the gapped bulk state; when approaching \(g_{c}\), the edge and bulk become gapless 
simultaneously, 
leading to the nonordinary surface transition \cite{zhanglong, Ding2018}.
However, this scenario was challenged by the finding of Weber and 
Wessel that the same dangling-chain edge of the $S=1$ model shows similar nonordinary SCBs. \cite{Weber2}

In a recent work on an improved classical 3D O(3) model \cite{Toldin},
Parisen Toldin found that the surface exponent 
$y_s$ is relatively small and argued that the small value of $y_s$ leads to a 
large region of surface coupling, where effectively the scaling 
of SCBs is governed by the special fixed point, and the 
observed exponents are thus close to the special SCBs and this 
offers an understanding of the nonordinary SCBs of the $S=1$
quantum Heisenberg model. 
For the $S=1/2$ quantum Heisenberg model, Jian and coauthors argued that if the special transition occurs in the 
presence of VBS order, $\eta_\parallel$ should equal to 
the $S=1$ value, due to a direct magnetic-VBS transition\cite{CMJian}.
The nonperturbative effect due to the topological $\theta$ term to $\eta_\parallel$ is argued to be small \cite{ContinuousN}.

Meanwhile, it was proposed that the SPT 
state may lead to a gapless edge state even in the gapped bulk state, which, together 
with the bulk critical modes,
leads to nonordinary SCBs in the appropriate edge \cite{zhanglong}. 
This scenario is verified in coupled spin-1 Haldane chains. 
\cite{CHC}

In this section, we study the surface critical behavior on this dangling 
chain edge for the easy-plane XXZ models where the symmetry of the 
model is down to $U(1)$, and, therefore, the bulk
transition belongs to the 3D O(2) universality class. 
Considering the puzzling situation that occurs in the SU(2) models, we 
here study both the $S=1/2$ and the $S=1$ models.

\subsection{$S=1/2$ models}

\begin{figure}[ht!]
  \includegraphics[width=\columnwidth]{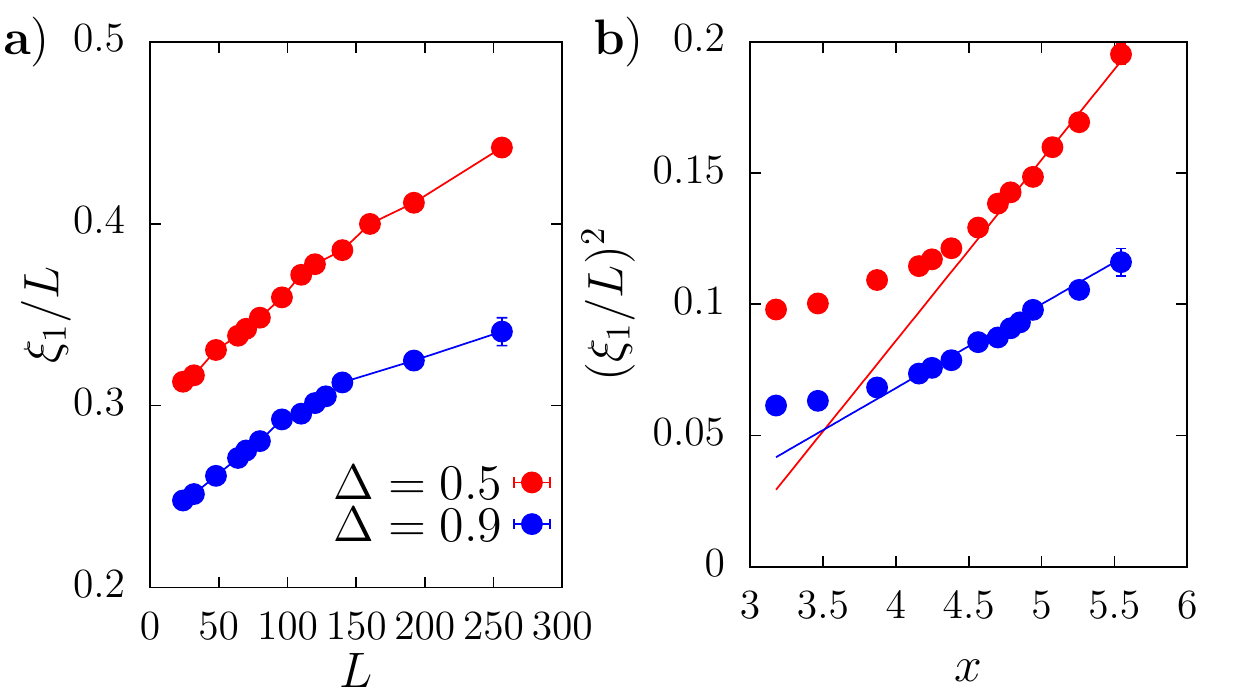}
    \includegraphics[width=\columnwidth]{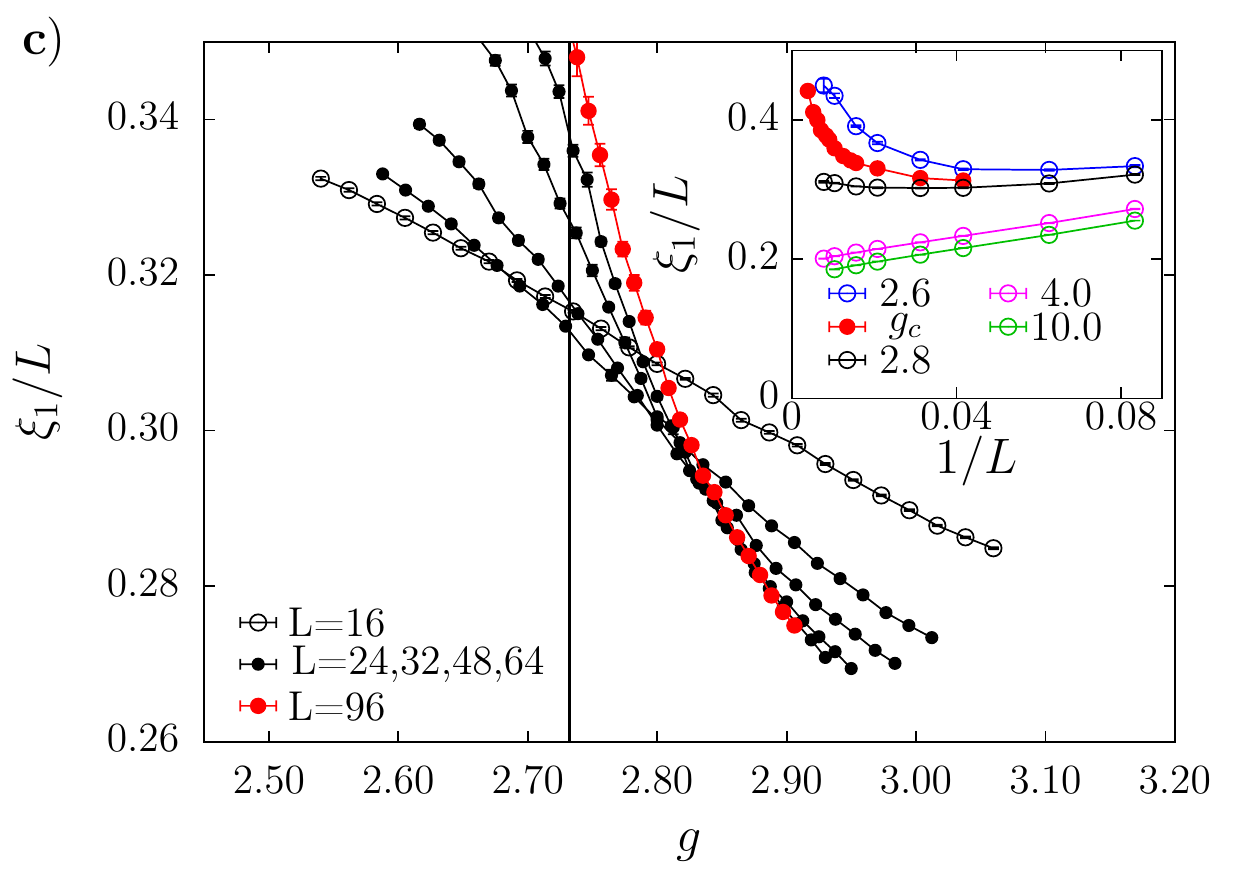}
  \caption{ a) The scaled surface correlation length \(\xi_1/L\) as a function of system size $L$ for the $S=1/2$ easy-plane CD models at
  bulk critical points $g_c$ on a lin-lin scale. b) 
  $(\xi_1/L)^2$ as a function of $x = \log(L/L_0)$ with $L_0$ constants different for $\Delta=0.5$ and $\Delta=0.9$.  
  For large enough $L$, the curves become linear.
    c) The scaled surface correlation length \( \xi_1/L \)  as a function
    of the bulk tuning parameter \(g\) for different 
  system sizes  for the \(S=1/2\) easy-plane CD model with \(\Delta=0.5\). The vertical line corresponds to the bulk
  critical point $g_c=2.73227$.
  We see crossings near  \( g_{cs} \approx 2.85\). 
  The inset plots \(\xi_1/L\) vs \(1/L\) for different \(g\). 
  \label{fig:xi1} }
  \end{figure}

Let us start with the $S=1/2$ models. Consider the  limit 
$g \to \infty$ where the dangling-chain edge 
is decoupled from the bulk, the edge should be a gapless and quasi-long-range 
ordered TLL \cite{Tomonaga, Luttinger, giamarchi}, where the vortex excitation is suppressed. 
If such gapless states continues to the 
bulk critical point, we expect nonordinary SCBs as those found in the 
SU(2) models, which can be considered the special transition point
of the surface transition induced by surface engineering \cite{Ding2018}.
As a result, the scaled surface correlation length $\xi_1/L$ should 
converge to a constant, different from 
the ordinary SCBs, where  $\xi/L$ goes to zero \cite{zhanglong}.

To test whether such behavior occurs or not, 
we calculated the scaled surface correlation length \(\xi_1/L\) 
on the dangling-chain edge for system sizes up to $L=256$ 
at the bulk critical points of $\Delta=0.5$ and
$\Delta=0.9$, respectively. 
To our surprise, we found that $\xi_1/L$  
diverges slowly with the system size $L$, as shown  
in Fig. \ref{fig:xi1}(a). 

To understand this observation, we then calculated $\xi_1/L$ as a function of 
$g=J'/J$ away from the bulk critical
point $g_c=2.73227$ for the $S=1/2$ $\Delta=0.5$ CD easy-plane XXZ model.
The results are plotted in Fig. \ref{fig:xi1}(c). 
Remarkably, we see the curves of $\xi_1/L$ vs. $g$
for different system sizes cross roughly at \(g_{cs} \approx 2.85\), larger than $g_c$. 

One possible scenario is that 
a phase transition occurs on the dangling-chain edge to precede the bulk transition.
This possiblity is unusual at first sight: 
inside the gapped dimer phase, before the bulk critical point is 
approached, the coupling between the dangling-chain edge and the bulk
can be considered weak. 
The Mermin-Wagner theorem \cite{Mermin} excludes the symmetry breaking  
phase in a (1+1)-D quantum system; therefore, no phase transition to an
ordered phase is allowed when $g$ is decreased.
Instead, one expects that  the $S=1/2$ dangling-chain edge stays
gapless or quasi-long-range 
ordered until the bulk critical point is reached, whereas critical 
fluctuations of the bulk may induce effective long-range interactions along 
the chain, and a phase transition to an ordered state may occur. 
However, the Mermin-Wagner theorem does not exclude phase transition
between two different short-range ordered phases \cite{Domany, WAGuo}.
Thus, a transition in principle does not violate any physical law. 
In addition, as found in classical models, the changing of interaction range, 
not necessarily infinite, can drive the phase to another fixed point\cite{xfqian}. 

To better understand the results and further explore the 
physics properties of the surface state, 
we calculated the surface correlation function \( C_\parallel(L/2) \) 
and $C_\perp(L/2)$, and the squared surface staggered magnetization $m_{s1}^2$. 

\begin{figure}
    \includegraphics[width=1\columnwidth]{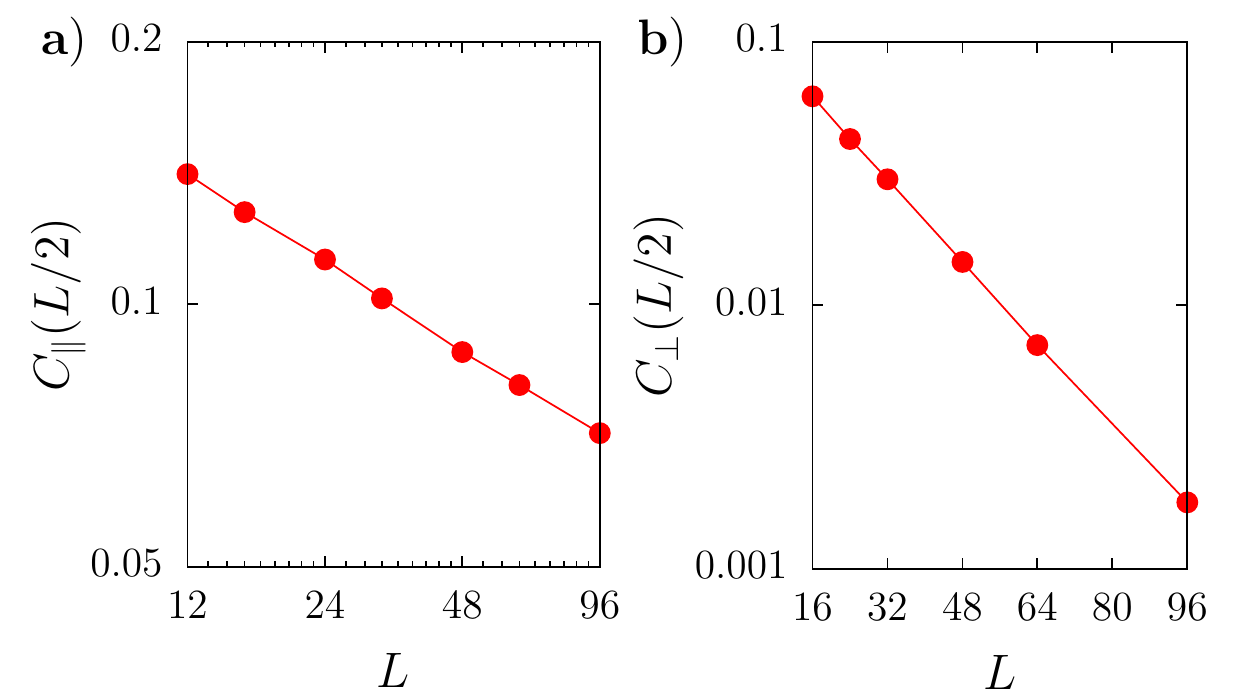}
    \caption{
        Surface correlations on the dangling-chain edge for the $S=1/2$ easy-plane CD model with $\Delta=0.5$ at $g=2.8$. 
        a) $C_\parallel(L/2)$ vs $L$ on a log-log scale. 
        b) $C_\perp(L/2)$ vs $L$ on a lin-log scale. 
        \label{fig:Cc2d05dis}
    }
\end{figure}

We first study the surface state at $g=2.8$ between the bulk critical point
$g_c=2.73227$ and the surface transition point $g_{cs} \approx 2.85$, 
where the bulk is in a gapped disordered phase of 
the $S=1/2$ CD XXZ model with $\Delta=0.5$. 
We find that $C_\perp(L/2)$ shows an exponential decay, as illustrated
in Fig. \ref{fig:Cc2d05dis}(b).
Fitting $C_\perp(L)$ according to the following equation
\be
C_\perp (L) = a {\rm e}^{-L/\xi_\perp},
\label{Cperpexp}
\ee
we obtain $\xi_\perp=22.4(2)$. 
It is evident that the bulk is gapped from the surface state at $g=2.8$. As a result, we 
do not expect long-range order on the surface. 

To confirm that there is no long-range order on the (1+1)-D dangling-chain edge, we fit $C_\parallel(L/2)$ according to the following scaling 
form:
\be
C_\parallel (L/2) = C_\parallel + a  L^{-(d+z-2+\eta_\parallel)},
\label{eqn:cpaext}
\ee
where $C_\parallel \ne 0$ gives the long-range order parameter, and $a$ is an unknown constant. 
The data of $C_\parallel(L/2)$ vs. $L$ are plotted in Fig. \ref{fig:Cc2d05dis}(a) on a log-log scale.
We find $C_\parallel=0.013(7)$, and $\eta_\parallel=-0.62(3)$. The fit is 
good, but the magnitude of $C_\parallel$ is within two times the error bar. We 
conclude that it is actually zero.

We then tried to fit $C_\parallel(L/2)$ according to Eq. (\ref{cpara}).
We find statistically sound fits for $L \ge 12$ 
without the $\omega$ term and obtain $\eta_\parallel = -0.670(5)$. 
This fitting result is stable under further exclusion of small system sizes.

\begin{figure}[h]
   \includegraphics[width=\columnwidth]{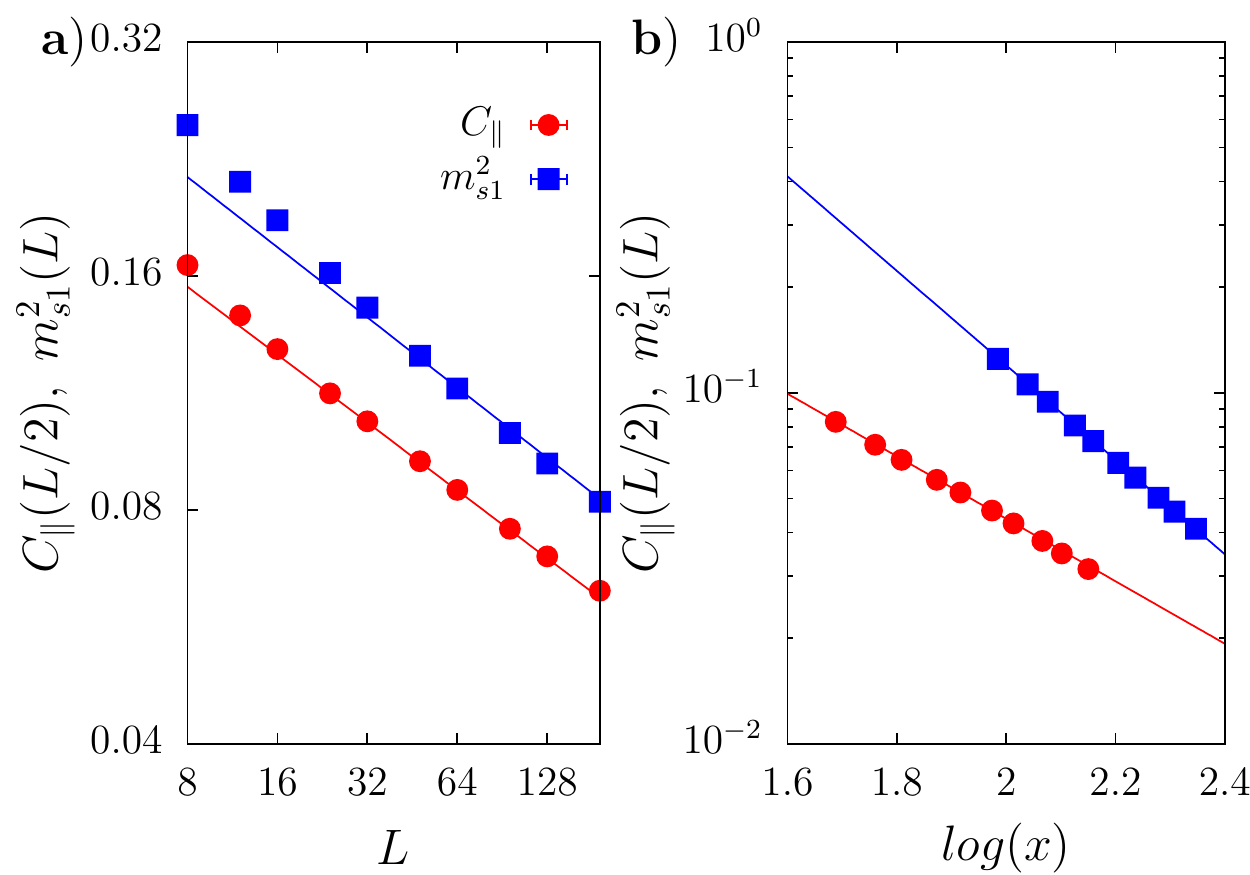}
\caption{ The surface behaviors of the $S=1/2$ easy-plane CD model with $\Delta=0.5$ at $g=2.77$.
a) The surface squared antiferro-magnetization
\(m^2_{s1}\) and the surface correlation function \(C_\parallel(L/2)\) vs 
$1/L$ . b) $C_\parallel(L/2)$ and $m^2_{s1}(L)$ vs $x=\log(L/L_0)$ on a log-log 
scale with $L_0=0.036(15)$ for $C_\parallel(L/2)$ and $L_0 = 0.006(1)$ for $m^2_{s1}(L)$. The lines are fitting curves.
  \label{fig:Ex77}}
\end{figure}

However, the power-law decay of the spin-spin correlation cannot support 
the divergence of $\xi_1/L$ found in the region between $g_c$ and $g_{cs}$, see Fig. \ref{fig:xi1}(c). Inspired by the recently 
proposed extraordinary-log phase at the bulk critical point\cite{ContinuousN}, we propose, in this phase, 
the correlation function $C_\parallel(L)$ decays as a power of $\log (L/L_0)$, i.e.,
\be
C_\parallel(L) \propto \log (L/L_0)^{-q} 
\label{Cplog}
\ee
where $q$ is an unknown exponent. At $g=2.77$, slightly closer to $g_c$, we calculated $C_\parallel$ and $m_{s1}^2$ up to $L=192$. The results
are shown in Fig. \ref{fig:Ex77}.
We fit $C_\parallel(L/2)$  and $m_{s1}(L)$ according to Eq. (\ref{Cplog}). 
We find statistically sound fit with $L_{min} \ge 24$ for both quantities, as illustrated in Fig. \ref{fig:Ex77}(b).  
However, the fitting results are not stable as more system sizes are excluded.
Instead, fitting  $C_\parallel(L/2)$ according to Eq. (\ref{cpara}) gives statistically sound results for $L_{min} \ge 16$, 
which are stable upon further excluding systems sizes.
Similarly, fitting $m_{s1}^2(L)$ according to Eq. (\ref{ms1fs}) offers good results for $L_{min} \ge 24$ and the fits are stable for further 
excluding small system sizes.
The fits are shown in Fig. \ref{fig:Ex77}(a).

Although we cannot exclude the possibility that there exists a surface phase transition preceding the bulk transition, 
our data do not support this scenario. 
Since the crossing points are rather close to the bulk critical point where the bulk correlation length is large, 
we turn to believe that
the divergence of $\xi_1/L$ at $g>g_c$ are results of the bulk critical behavior. 
It is more likely
that the correlation behaves finally cross over to normal TLL behavior.

\begin{figure}[h]
   \includegraphics[width=\columnwidth]{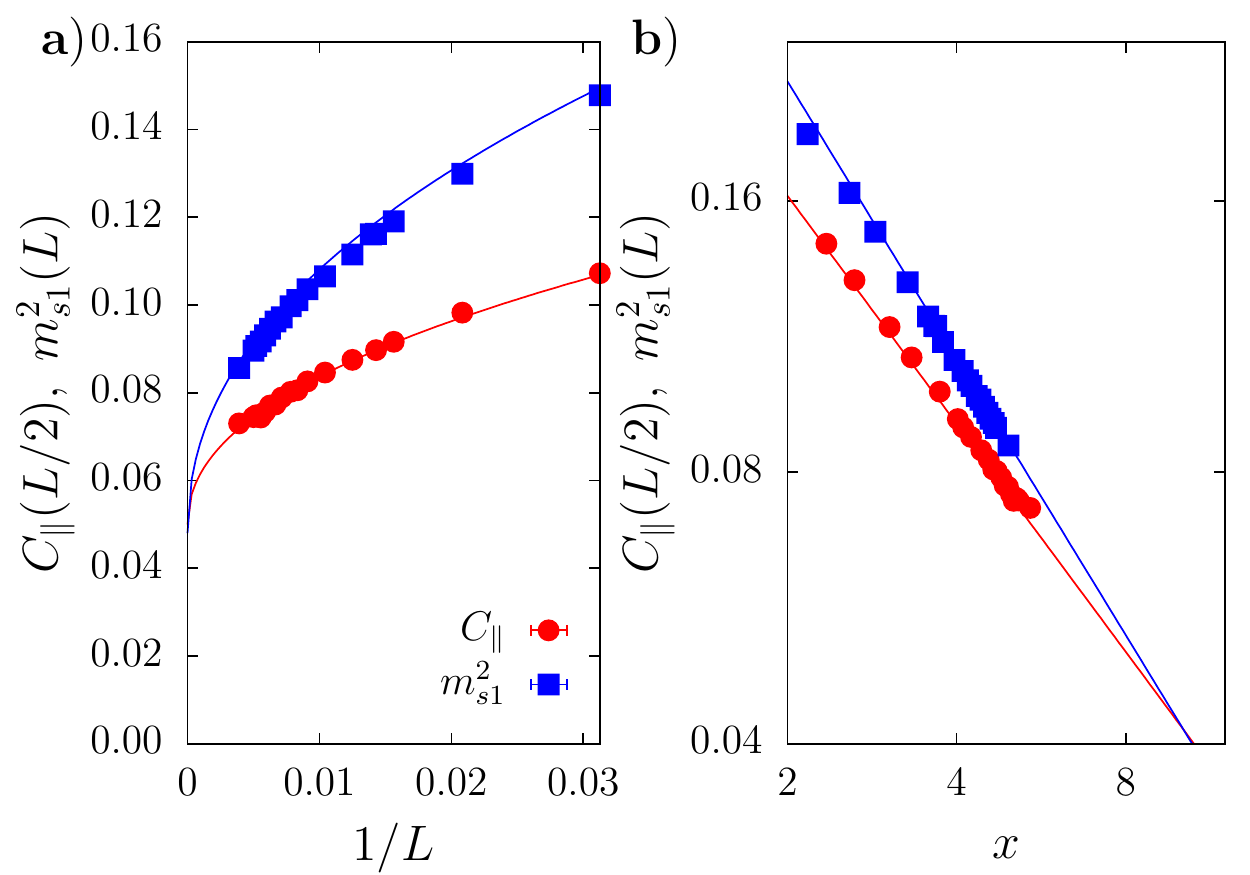}
\caption{ The surface behaviors of the $S=1/2$ easy-plane CD model with $\Delta=0.5$ at the bulk critical point.
a) The surface squared antiferro-magnetization
\(m^2_{s1}\) and the surface correlation function \(C_\parallel(L/2)\) vs 
$1/L$ . b) $C_\parallel(L/2)$ and $m^2_{s1}(L)$ vs $x=\log(L/L_0)$ on a log-log 
scale with $L_0=1.2(1)$ for $C_\parallel(L/2)$ and $L_0 = 1.8(7)$ for $m^2_{s1}(L)$. The lines are fitting curves.
  \label{fig:Ex}}
\end{figure}

We now move to the bulk critical point $g_c$.

In general, the scaling behavior in Eq. (\ref{eqn:cpaext}) is expected for extraordinary SCBs where the surface orders 
preceding the 
bulk when $g$ is decreased, 
and exhibits extra singularities at the bulk critical point, for example,  
at the bulk critical point of a classical O($n$) model in dimensions $d>3$ with large surface couplings.
Furthermore, according to Eq. (\ref{eqn:mxy}), we
expect similar behavior of $m_{s1}^2$
\be
m_{s1}^2 (L)= m^2_{s1} + b L^{2y_{h1}-2 (d+z-1)}
\label{eqn:m1ext}
\ee
with magnetization $m^2_{s1}\approx C_\parallel$ and an unknown constant $b$.
The results of $C_\parallel (L/2) $ and $m_{s1}^2(L)$ vs. $1/L$ at the bulk critical 
point $g_c$ for the $S=1/2$  
CD easy-plane XXZ model with $\Delta=0.5$ are shown in Fig. \ref{fig:Ex}(a). 

However, for a classical O(2) model in three dimensions, the conclusion is controversial. 
At sufficiently large surface couplings, the surface should undergo a 
BKT transition to a quasi-long-ranged order 
before the bulk orders when the temperature is lowered down. When the 
bulk critical point is approached, Deng {\it et al.} proposed that the surface
becomes long-range ordered due to long-range interactions induced by
critical fluctuations; therefore, the order parameter has a jump\cite{Deng}. As a 
result, the surface magnetization 
$m_1^2$ follows the scaling form Eq. (\ref{eqn:m1ext}) with
$m_{s1}^2(L)$ replaced by $m_1^2(L)$.
The nonzero $m_{1}^2$ and the exponent $y_{h1}$ 
describing the power-law decay of the surface magnetization
are found numerically \cite{Deng}. However, Deng {\it et al.} offered another
possibility that the surface magnetization still 
decays to zero in a power-law\cite{Deng}. Moreover, a slightly different
exponent $y_{h1}$ was found by using the same data.
For the reader's convenience, we list both results in Table \ref{tab:ext}.

If the former scenario is realized on the dangling-chain surface 
in our QAF easy-plane model, we expect 
the surface state to lie in 
an extraordinary state similar to that of the classical O(n) model
with dimension $d>3$, 
which means $C_\parallel(L)$ behaves as Eq. (\ref{eqn:cpaext})
with $C_\parallel \neq 0$. 

\begin{table*}[ht] 
   \caption{ Surface critical properties on the dangling-chain edge of the easy-plane CD models with $S=1/2$ and $S=1$. The SCBs of the 
   classical 3D O(2) model at large surface coupling are also listed here for comparison. 
   Here, $\eta_\parallel$ are calculated using the scaling law $\eta_\parallel=3-2y_{h1}$. }
    \centering
    \begin{tabularx}{\textwidth }{c @{\extracolsep{\fill}}c c c c c c c c  }
    \toprule
     Type & $\Delta$ &$C_\parallel$& $m^2_{s1}(m_1^2)$ & $y^{(1)}_{h1}$ & $\eta_\parallel$ & $\eta_\perp$ 
      &$q$ & $\alpha$\\
       \hline
       \multirow{2}{*}{$S=1/2$}
    &0.5 &0.049(2) & 0.048(4)         &  1.77(2)          &-0.55(2) & -0.415(2)&  0.82(2)& 0.22(3)\\
    &0.9 &0.033(2) & 0.034(2)         &  1.69(1)         &-0.41(2) & -0.43(1)& 1.03(2) & 0.10(1)\\
     \hline
     \multirow{1}{*}{$S=1$}
     &0.9 &0.31(1)   & 0.34(2)             & 1.76(2) &-0.55(2) &   -0.418(2)   & 0.6(1)   & 0.13(7) \\ 
    \hline
          \multirow{2}{*}{Classical XY}&
          \multirow{2}{*}{0}
     & $-$ & 0.222\cite{Deng} 
     &1.812(5) \cite{Deng}  &-0.62(1) &       &  0.59(2)\cite{lvjp} & 0.27(2) \cite{lvjp}\\
     &  & $-$    & $-$    
     &1.9675(30)\cite{Deng} &-0.935(6) &         &        &\\
     \hline
       \toprule
    \end{tabularx}
    \label{tab:ext}
\end{table*}

We tried to fit Eqs. (\ref{eqn:cpaext}) and (\ref{eqn:m1ext}) to our data.
By gradually discarding small system sizes, we obtained a statistically sound fit 
without correction to scaling term.
From correlation data $C_\parallel(L)$, we find asymptotic value 
\(C_\parallel =0.049(2)\) and $\eta_\parallel=-0.55(2)$, for $\Delta=0.5$.  
For $\Delta=0.9$, we estimated $C_\parallel=0.033(2)$, and $\eta_\parallel=-0.41(2)$. 
By fitting the magnetization data, we find \(m^2_{s1}=0.048(4)\)
and $y_{h1}=1.77(2)$ for $\Delta=0.5$; \(m^2_{s1}=0.034(2)\)
and $y_{h1}=1.69(1)$ for $\Delta=0.9$. The
fitting functions for $\Delta=0.5$ are graphed in Fig. \ref{fig:Ex}(a).
The above results are listed in Table \ref{tab:ext}.
The estimated $\eta_\parallel$ and $y_{h1}$ for both $\Delta$ satisfy the scaling relation Eq. (\ref{scaling2}).
The two constants $C_\parallel$ and $m_{s1}^2$ are also in good agreement.

Though the values of $C_\parallel$ and $m_{s1}^2$ found by our fitting 
are very small, their magnitudes are much larger than the corresponding error bars.  It is also possible to fit without
$C_\parallel$ and $m_{s1}^2$, but both fits to \(C_\parallel(L/2)\) and \(m^2_{s1}(L) \) give much worse fitting 
performance, even when the corrections to scaling are included. We therefore 
exclude this possibility.

We have also calculated the $C_\perp(L/2)$ at $g_c$. The 
correlation $C_\perp(L/2)$ of the 
extraordinary SCBs should scale as Eq. (\ref{cperp}), but with 
a different $\eta_\perp$ from that of the ordinary SCBs. 
By fitting according to Eq. (\ref{cperp}),
we find $\eta_\perp=-0.415(2)$ for $\Delta=0.5$ and $\eta_\perp=-0.43(1)$ for 
$\Delta=0.9$. These results are also listed
in Table \ref{tab:ext}.   Indeed, they are different from 
$\eta_\perp$ of the dangling-ladder edge.

Recently, Metlitski \cite{ContinuousN} proposed a theory based on
RG arguments that an extraordinary-log phase may occur on the surface
of the classical 3D O(2) model when the surface couplings are strong enough.
The quasi-long-range ordered surface phase undergoes a phase transition to
the extraordinary-log phase, while stiffness of the surface order parameter diverges logarithmically.
Such a transition is different from the suggestion proposed in \cite{Deng} that 
a first-order
transition with a jump of order parameter occurs when the bulk critical
point approaches from the quasi-long-range ordered surface phase.
The extraordinary-log state is a state less ordered than the long-range 
ordered state but more ordered than the quasi-long-range ordered state, 
in which the correlation function decays as a power of the logarithm of 
distance \cite{ContinuousN}.
Numerical works on classical 3D O(3)  and 3D O(2) models 
support the existence of such a phase \cite{Toldin,lvjp}.

We then explored the possibility that the surface state at $g_c$ is  
the extraordinary-log state.  As proposed in [\onlinecite{ContinuousN}], in 
such a state, the correlation function $C_\parallel(L/2)$ decays as a power of $\log (L/L_0)$, in the form of Eq. (\ref{Cplog}).
Such behavior has been observed with 
$q=0.59(2)$ on the surface of the critical 3D classical O(2) model with sufficiently
enhanced surface couplings \cite{lvjp}; see also Table \ref{tab:ext}.
We replot the surface parallel correlation function $C_\parallel(L/2)$ on the 
dangling-chain surface for the $S=1/2$ CD easy-plan models in
Fig. \ref{fig:Ex}(b) on a log-log scale with 
$x=\log(L/L_0)$. We observe a good straight curve.
By gradually excluding small system sizes, we obtain statistically sound
fits of the data according to Eq. (\ref{Cplog}). 
The best fits show $q=0.82 (2)$ and $L_0=1.2(1)$ for $\Delta=0.5$, and $q=1.03 (2)$ and 
$L_0=1.26(9)$ for $\Delta=0.9$ in the CD model.  These
results are listed in Table \ref{tab:ext}. The fitting curve for the 
$\Delta=0.5$ model is graphed in Fig. \ref{fig:Ex}(b).

Furthermore, the extraordinary-log scaling offers another scenario to understand the divergence of $\xi_1/L$:
according to the extraordinary-log scaling, the TLL parameter diverges logarithmically with the bulk
correlation length as one approaches $g_c$\cite{Maxprivate}, leading to the divergence of $\xi_1/L$ near $g_c$.

The finite-size behavior Eq. (\ref{Cplog}) suggests that the surface magnetization $m_{s1}(L)$ also decays logarithmically with the same exponent $q$.
We then tried to fit this formula to our $m_{s1}^2(L)$ data. However, the
fits for both the $\Delta=0.5$ and $\Delta=0.9$ $S=1/2$ models were not 
stable when we excluded small system sizes gradually. The asymptotic behaviors
of $C_\parallel(L/2)$ and $m_{s1}^2(L)$ were also very different, as shown in 
Fig. \ref{fig:Ex}(b) for $\Delta=0.5$.

In addition, in recent work on the SCBs of an improved model in the classical 
3D O(3) universality class \cite{Toldin}, where the extraordinary-log phase is 
verified at strong surface couplings, it is further proposed that $\xi_1/L$ scales as
\be
(\xi_1/L)^2 = \frac{\alpha}{\pi(n-1)}\log (L/L_0)
\label{xi1log}
\ee
with $\alpha$ a constant for the O($n$) model and $\pi$ represents the antiferromagnetic momentum. Here, we try to fit this ansatz 
to our $\xi_1/L$ data. We again find statistically sound fits when small system 
sizes are excluded.
The results are, as listed in Table \ref{tab:ext}, $\alpha=0.22(3), L_0=16(8)$ 
for $\Delta=0.5$ and $\alpha=0.10(1), L_0=7(3)$ for $\Delta=0.9$, respectively.
These results are demonstrated by replotting the rescaled 
surface correlation length $(\xi_1/L)^2$ as functions of $x=\log(L/L_0)$, 
with $L_0$ set to the fitting results, in Fig. \ref{fig:xi1}(b). 
The curves are only linear for sufficiently large system sizes. 

However, the values of $\alpha$ and $q$ estimated here do not agree with
the corresponding values of the classical O(2) model found 
in \cite{lvjp}, and they do not satisfy the scaling relation
\begin{equation}
    q=\frac{n-1}{2\pi \alpha},
    \label{eqn:qalpha}
\end{equation}
predicted in \cite{ContinuousN}, with $n$ being the number of components of 
the classical spin. 

Metlitski also proposed a possible state in the extraordinary-log
fixed point with coexisting charge-density-wave order for the quantum O(2) model.
We have calculated the surface correlation 
$C_\parallel^{\rm Z}(L/2)$ that averages $C^{\rm Z}(i,j)$ between two surface
spins $i$ and $j$ at the longest distance $L/2$ over the dangling chain. We find that 
the correlation goes to
zero as $L \to \infty$. Thus, no charge-density-wave order is found.

Considering all these analyses, for the $S=1/2$ easy-plane XXZ model, 
we conclude that our numerical results support an extraordinary transition on the dangling-chain surface with a 
long-range order established 
by effective long-range interactions due to bulk critical fluctuations. Moreover this 
suggests the possibility that the state is an extraordinary-log state seems unlikely. The extraordinary-log scenario is also excluded for the SU(2) CD model.\cite{ding2021}

\begin{figure}
    \centering
    \includegraphics[width=1\columnwidth]{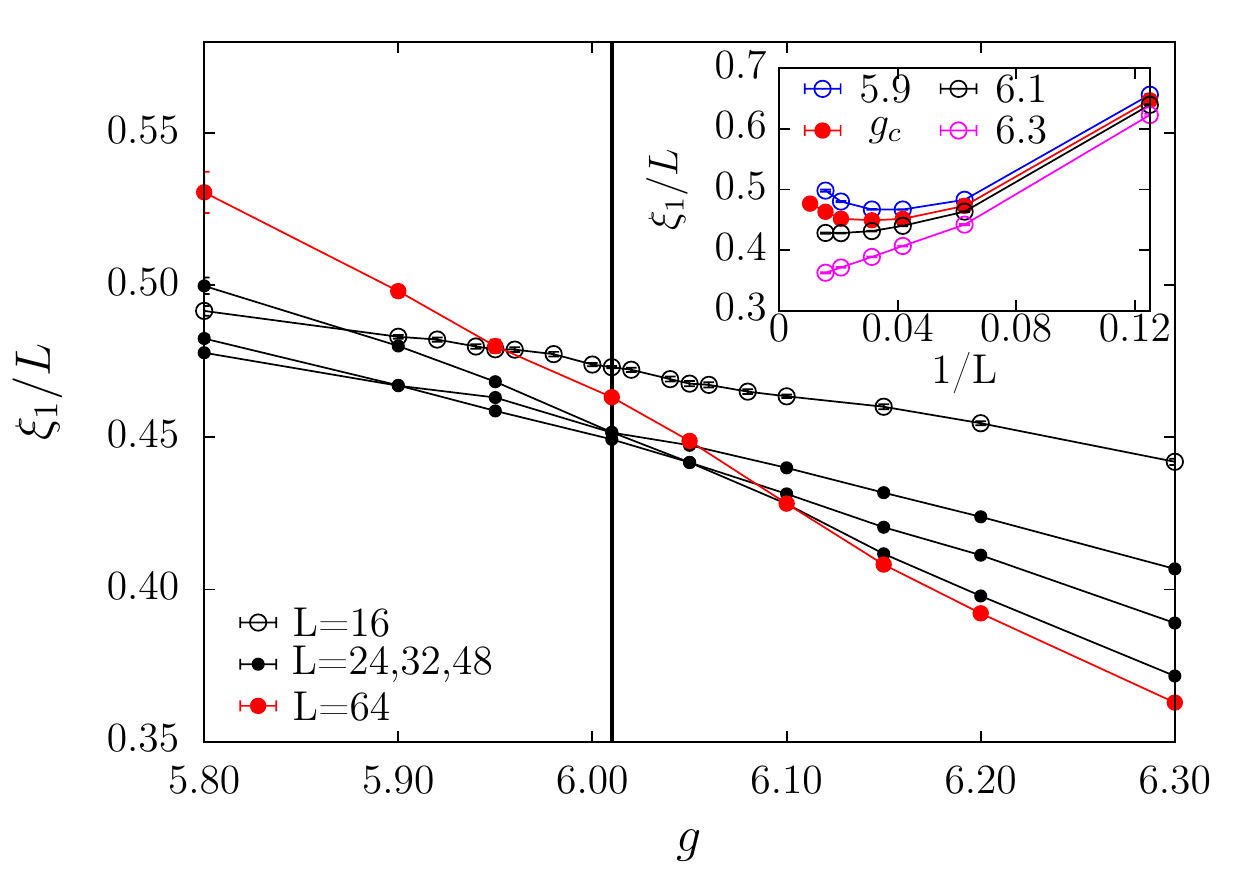}
    \caption{The scaled surface correlation length $\xi_1/L$ on the dangling-chain edge of the $S=1$ easy-plane CD model with $\Delta=0.9$ 
    vs $g$ for different system sizes $L$, showing a surface transition in the bulk gapped phase. The vertical line corresponds to the bulk critical point $g_c=6.01$. The inset plots $\xi_1/L$ vs $1/L$ for different $g$, showing divergence of $\xi_1/L$ at $g_c$.}
    \label{fig:S1xi1g}
\end{figure}

\subsection{$S=1$ model}

To further investigate the nature of the dangling-chain surface, we also 
performed simulations for the $S=1$ CD easy-plane model with $\Delta =0.9$. 

We find a surface transition in the bulk gapped phase, 
when $g$ is adapted, by calculating the dimensionless ratio
$\xi_1/L$ as a function of $g$ for different system sizes. As shown in 
Fig. \ref{fig:S1xi1g}, the curves of $\xi_1/L$ show crossings that converge to a
value around $g_{cs} \simeq 6.1$, larger than the bulk critical point $g_c=6.01$ for 
the $S=1$ CD easy-plane model with $\Delta=0.9$, in the gapped bulk phase, 
indicating a phase transition. 
Different from the $S=1/2$ case, the edge is an $S=1$ 
chain  with a Haldane gap at the large $g$ limit; therefore,
this transition can be understood as a BKT transition in 
a (1+1)-D system. Without topological terms and thanks to the dangling settings, 
the system is analogous to the 3D classical O(2) model with strong surface couplings, which lead 
to the BKT transition in the 
bulk disordered phase when the temperature is lowered. Indeed, $C_\parallel(L/2)$ shows
a power-law decay $1/r^x$ at $g=6.1$. Fitting for $L$ in the range from 16 to 64, we find $x=0.22(2)$, which
is close to 1/4 at the BKT transition. 

\begin{figure}[h]
   \includegraphics[width=\columnwidth]{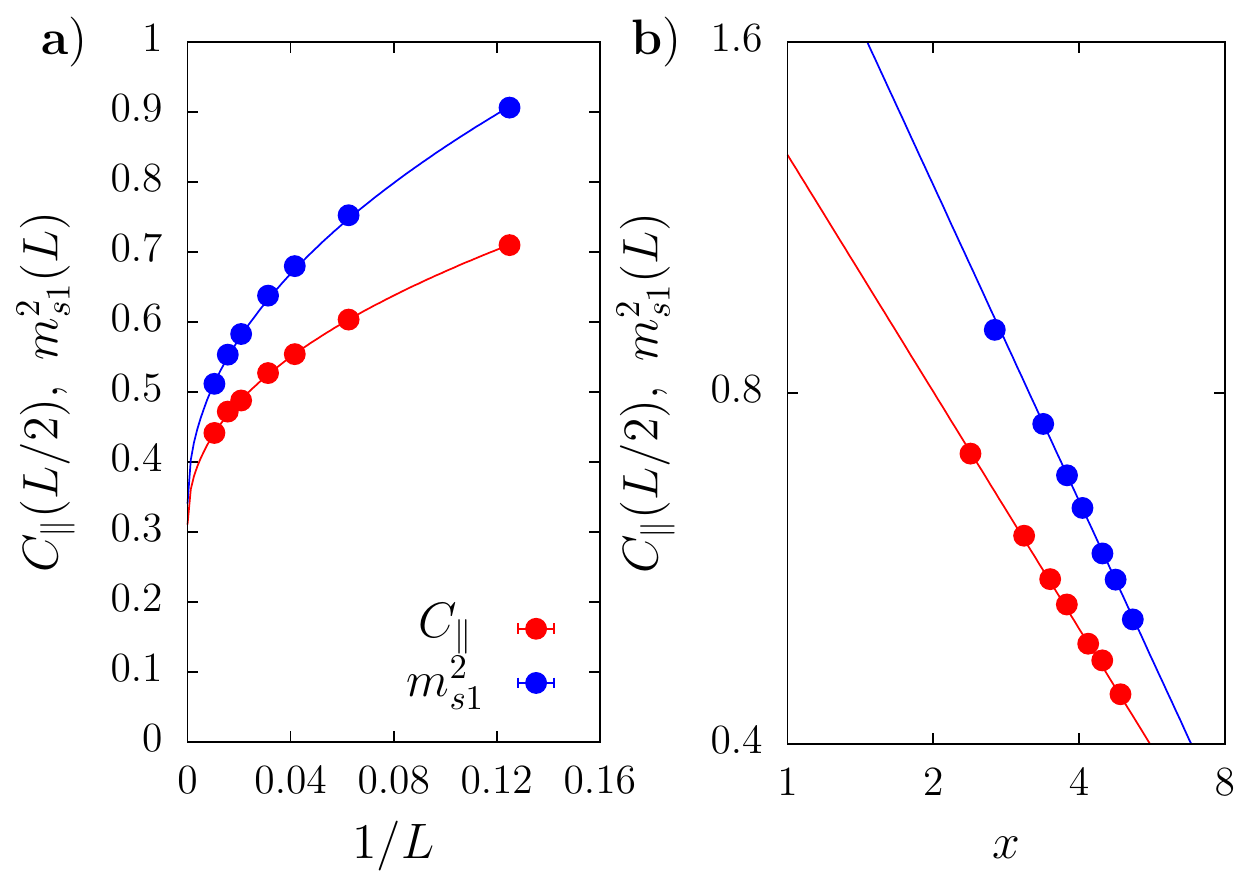}
\caption{ The surface behaviors of the $S=1$ easy-plane CD model with $\Delta=0.9$ at the bulk critical point.
a) The surface squared antiferro-magnetization
\(m^2_{s1}\) and the surface correlation function \(C_\parallel(L/2)\) vs 
$1/L$ . b) $C_\parallel(L/2)$ and $m^2_{s1}(L)$ vs $x=\log(L/L_0)$ on a log-log 
scale with $L_0=1.5(1.0)$ for $C_\parallel(L/2)$ and $L_0 = 0.5(4)$ for $m^2_{s1}(L)$. The lines are fitting curves.
  \label{fig:S1Extr}}
\end{figure}

We then studied the surface critical behavior at the bulk critical 
point $g_c=6.01$. The results of $C_\parallel(L/2)$ and $m_{s1}^2(L)$ vs. $1/L$ at $g_c$
are shown in Fig. \ref{fig:S1Extr}(a).
Since $g_c<g_{cs}$ and no topological terms available for the $S=1$ chain and dangling
settings, we expect a
similar SCB to that in the classical 3D O(2) model with surface coupling enhanced. 
We therefore try to fit the results of $C_\parallel(L/2)$ 
using Eq. (\ref{eqn:cpaext}) and $m_{s1}^2(L)$ using Eq. (\ref{eqn:m1ext}) as 
expected for the extraordinary SCBs of the classical 3D O(2) model, where long-range 
order is induced by effective long-range couplings due to critical modes, as 
proposed in \cite{Deng}. 

From data of $C_\parallel(L/2)$,  we find $C_\parallel=0.31(2)$, 
and $\eta_\parallel = -0.55(2)$, for system sizes
in the region $L\in [8,64]$. The fit is statistically sound. 
By fitting Eq. (\ref{eqn:m1ext}) to $m_{s1}^2(L)$, we find $m^2_{s1}=0.34(2)$ and $y_{h1}=1.76(2)$. These results are listed in Table \ref{tab:ext}.

It is somewhat surprising that the values of $C_\parallel$ and $m^2_{s1}$ are
much larger than those of the $S=1/2$ case.
Semiclassically, the length of the $S=1$
spin is $\sqrt{2}$, and the length of the $S=1/2$ is 
$\sqrt{3}/2$. In this sense, if the models are in the same
classical states, this value of $C_\parallel$ (for $S=1$) is too large and cannot 
be simply attributed to the size of the spin vector.
This suggests that the quantum character of the  $S=1/2$ model is very important.
However, the exponents $\eta_\parallel$ and $y_{h1}$ for the $S=1$ model are close 
to those 
of the $S=1/2$ SCB. 

The behavior of $C_\perp(L/2)$ apparently follows the standard
scaling behavior Eq. (\ref{cperp}). 
From data of $C_\perp(L/2)$, we find statistically sound fit in 
the region  $L\in [8,64]$. 
The fitting results, listed in Table \ref{tab:ext}, are 
$\eta_\perp = -0.418(2)$.

Considering recent progress in the SCBs of the 3D O(2) model \cite{lvjp},
it is tempting to fit Eq. (\ref{Cplog}) to the $C_\parallel(L/2)$ data 
and fit Eq. (\ref{xi1log}) to the $\xi_1/L$ data of 
the $S=1$ model.
We find a statistically sound fit with $q=0.6(1)$ and $L_0 = 1.5(1.0)$ for Eq. (\ref{Cplog}). The data and fitting curves are plotted on a log-log scale in
Fig. \ref{fig:S1Extr}(b).
The constants $\alpha=0.13(7)$ and $L_0=0.3(6)$ in Eq. (\ref{xi1log}) were estimated. 
All the results are summarized in Table \ref{tab:ext}.
It is interesting that the exponent $q$ equals the value of the classical 
model with $q=0.59(2)$. However, the value of $\alpha$ is different from the 
corresponding value of the classical model, and the relation Eq. (\ref{eqn:qalpha})
is not satisfied.
In addition, we have also fit $m_{s1}^2(L)$ data with Eq. (\ref{Cplog}) and found statistical sound fit $q=0.9(2)$ and $L_0=0.5(4)$, as shown in Fig. \ref{fig:S1Extr}.

While we cannot exclude the scenario in which the SCB of the dangling chain
edge of the $S=1$ CD model belongs to the extraordinary fixed point for
normal O(n) model, especially when considering the large nonzero values of $C_\parallel$ and $m_{s1}^2$, however, from the point of quantum and classical mapping, it 
is likely that the SCBs are of the new extraordinary-log type.

\section{Conclusion}
\label{conclusion}

Using unbiased quantum Monte Carlo simulations, we have studied 
the 2D columnar dimerized quantum antiferromagnetic
XXZ model with easy-plane anisotropy for both $S=1/2$ and $S=1$ spins. We have paid 
particular attention to the surface behaviors on two different surfaces exposed
by cutting a row of strong bonds or cutting a row of weak bonds along the 
direction perpendicular to the strong bonds of a periodic system.

We found ordinary SCBs on the dangling-ladder surface for both $S=1/2$ and $S=1$ systems, 
but completely different surface behaviors on the dangling-chain surface for the $S=1/2$ and $S=1$ models. 

On the dangling-chain surface, the $S=1$ model behaves akin to the 
classical 3D XY model with large surface couplings. We find a BKT transition in the 
bulk gapped phase, and the surface seems to show extraordinary-log
behaviors at the bulk critical point.

For the $S=1/2$ model, we find unexpected crossings of scaled surface correlation length $\xi_1/L$
in the bulk gapped phase, considering the surface is already critical as 
a TLL at the large $g$ limit. One scenario to explain this is that there is a new surface state,
between the transition point and the bulk critical point, 
with spin-spin correlation $C_\parallel(L/2)$ decaying as a power of $\log(L)$. As a result, $\xi_1/L$
diverges in this phase, as well as at the bulk critical point.
However, fitting of numerical data upto $L=192$ does not support this scenario, the fits are not stable
upon gradually excluding small system sizes. The correlation behaves seem cross over to normal TLL power
law behavior, as small system sizes are excluded in the fits. 
In the end, it is more plausible that the divergence of $\xi_1/L$ are results of bulk critical behavior.

At the bulk critical point, our numerical 
results support an extraordinary SCB with a long-range 
order established by effective long-range interactions due to bulk critical 
correlations. The possibility that the surface shows 
extraordinary-log behavior seems unlikely. This is very different from the 
SU(2) dimerized Heisenberg model, 
where the surface transition at the dangling-chain edge is found 
to be nonordinary, but also not extraordinary\cite{Ding2018,Weber1}, in the sense 
that $C_\parallel$ and $m_{s1}^2$ were found to converge to zero.   

It would be very interesting to further investigate the phase transition from
TLL  to the critical state in the bulk gapped phase.

\begin{acknowledgments}
We thank Max Metlitski for pointing out the correct form of Eq. (\ref{xi1log}).	
W.Z. and W.G. were supported by the National Natural Science Foundation of China under Grant No.~11775021 and No.~11734002. C.D. was supported by the National Natural Science Foundation of China under Grant No.~11975024, No.~11774002 and No.~62175001 and the Anhui Provincial Supporting Program
for Excellent Young Talents in Colleges and Universities under Grant No.~gxyqZD2019023. 
L.Z. was supported by the National Natural
Science Foundation of China under Grant No.~11804337 and No.~12174387, the Strategic Priority Research Program of CAS under Grant No.~XDB28000000 and the CAS Youth Innovation Promotion Association. 
The authors acknowledge support extended by the Super Computing Center of Beijing Normal University.
\end{acknowledgments}

\bibliography{exlog.bib}

\newpage

\appendix
\newpage
\section{Bulk Critical Points}
\label{cri_bulk}
We determine the bulk quantum critical points of the two-dimensional 
columnar dimerized quantum 
antiferromagnetic easy-plane XXZ models with anisotropic parameter $\Delta < 1$ 
using finite-size scaling analyses, based on the 
SSE QMC simulations of the models on $L \times L$ square lattices with periodic boundary conditions. 
We set the inverse temperature $\beta$ to $2L$ to obtain the properties 
of the zero-temperature quantum phase transition at the limit $L\to \infty$.

We measure the spin stiffness $\rho^{\alpha}_s$ along the
$\alpha$ direction 
which is related to the winding number fluctuations as follows:
\begin{equation}
\rho^{\alpha}_s = \frac {1}  {\beta } \langle W_\alpha^2 \rangle,
\end{equation}
where $\alpha=x, y$, and $W_x (W_y)$ is the winding number in the $x$ ($y$) direction, which is defined as
\be
W_\alpha=(N_\alpha^+-N_\alpha^-)/L
\ee
where $N_\alpha^+$ ($N_\alpha^-$) denotes the number of operators transporting spin in
the positive (negative)  $\alpha$ direction.

We take the following scaling ansatz, namely, the scaled spinstiffness $L\rho_s^\alpha$ is dimensionless to extrapolate quantum 
critical point $g_c$ and the correlation length exponent $\nu$
\begin{eqnarray}
  \label{rs_form}
  L \rho_s^\alpha (g,L) = f((g-g_c)L^{1/\nu},L^{-\omega}),
\end{eqnarray}
where $\nu$ is the exponent for correlation length and $\omega$ is 
the leading exponent for the corrections to scaling. 
  This is done by expanding the above equation to
the second order of $(g-g_c)L^{1/\nu}$
\begin{equation}
    \begin{split}
   L \rho_s^{\alpha}(g,L) &= c + a_1 (g-g_c)L^{1/\nu}
   + 
a_2 L^{-\omega_1}\\
&+ b_1 (g-g_c)^2 L^{2/\nu} \\
  &+ b_2 (g-g_c)L^{1/\nu-\omega_1} + b_3 L^{-2\omega_1}  
    \end{split}
    \label{Lrho_scaling}
\end{equation}
then fitting the above formula to the data obtained from simulations.

\begin{figure}[b]
  \includegraphics[width=1.0 \columnwidth]{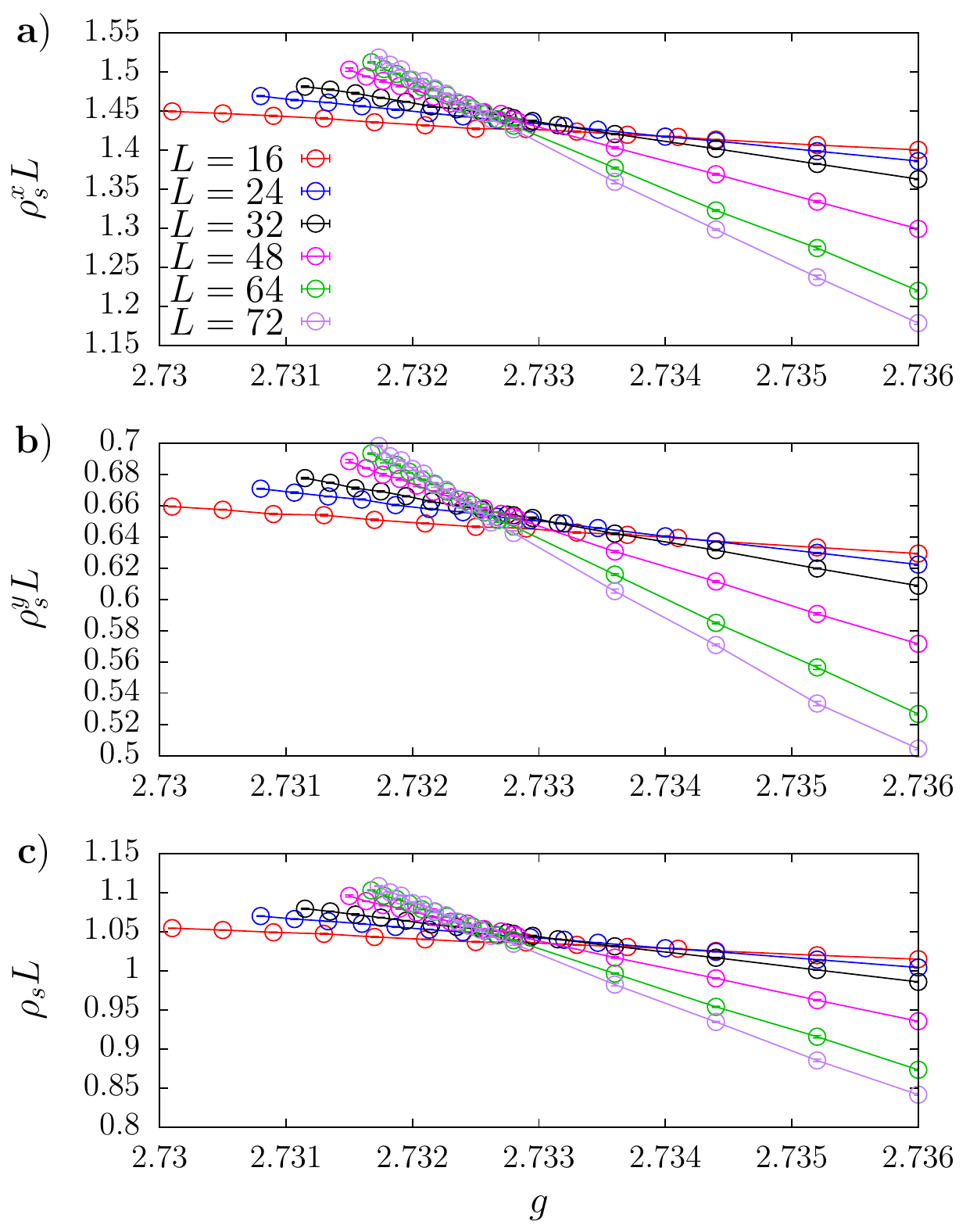}
  \caption{The scaled spin stiffness of the $S=1/2$ columnar 
  dimerized XXZ model with $\Delta=0.5$ for different system sizes $L$ as a function of g. a) $\rho_s^x L$ vs $g$, b) $\rho_s^y L$ vs $g$,
  c) $\rho_s L$ vs $g$.
  \label{fig:cri_C05}}
\end{figure}

Figure \ref{fig:cri_C05} shows the scaled spin stiffness $\rho_s^x L$, $\rho_s^y L$ and
$\rho_s L$ as a function of $g$ for various system sizes. 
We have done the fits for $\rho_s^x L, \rho_s^y L$ and $\rho_s L= \frac{1}{2}(\rho_s^x+\rho_s^y) L$. By gradually excluding data of small system sizes,
we obtain statistically sound and stable fits of Eq. (\ref{Lrho_scaling})
for $S=1/2$ models with $\Delta=0.5$ and $\Delta=0.9$, and $S=1$ model with $\Delta=0.9$.
The final estimates of $g_c$ and $\nu$ are listed in Table \ref{tab:cri}.

\end{document}